\documentclass[12pt]{iopart}
\usepackage{graphicx}

\begin{document}

\title{Comparison of 2D melting criteria in a colloidal system}

\author{Patrick Dillmann, Georg Maret, and  Peter Keim}

\address{Fachbereich Physik, Universit\"at Konstanz, 78457 Konstanz}
\ead{peter.keim@uni-konstanz.de}
\begin{abstract}
We use super-paramagnetic spherical particles which are arranged in a two-dimensional monolayer at a water/air
interface to investigate the crystal to liquid phase transition. According to the KTHNY theory a crystal melts in thermal equilibrium by two continuous phase transitions into the isotropic liquid state with an intermediate phase, commonly known as hexatic phase. We verify the significance of several criteria based on dynamical and structural properties to identify the crystal - hexatic and hexatic - isotropic liquid phase transition for the same experimental data of the given setup. Those criteria are the bond orientational correlation function, the Larson-Grier criterion, 2D dynamic Lindemann parameter, the bond-orientational susceptibility, the 2D Hansen-Verlet rule, the L\"{o}wen-Palberg-Simon criterion as well as a criterion based on the shape factor of Voronoi cells and  Minkowski functionals. For our system with long range repulsion, the bond order correlation function and bond order susceptibility works best to identify the hexatic - isotropic liquid transition and the 2D dynamic Lindemann parameter identifies unambiguously the hexatic - crystalline transition.

\end{abstract}

\vspace{2pc}
\submitto{\JPCM}

\section{Introduction}
\label{secintro}

While the liquid to crystal transition in three dimensional systems is usually a first order transition the situation in two dimensional systems is found to be more complex. While grain-boundary induced melting \cite{Chui1983,Kleinert1983} or condensation of geometrical defects \cite{Glaser1993,Lansac2006} suggest a first order phase transition, the theory of Kosterlitz, Thouless, Halperin, Nelson, and Young \cite{Kosterlitz1973,Nelson1979,Halperin1978,Young1979} predicted a melting process via two continuous phase transitions with an intermediate phase. The intermediate phase appears due to the fact that the translational and orientational symmetries are broken at different temperatures. The first phase transition at temperature $T_m$ is associated with destroying the discrete translational symmetry. The intermediate phase is named hexatic based on the remaining sixfold quasi long range orientational order. If the orientational symmetry is destroyed to a short range one at temperature $T_i > T_m$ a second phase transition from the hexatic to the isotropic liquid occurs. According to the KTHNY theory the different symmetries are affiliated with the occurrence of different topological defects, namely dislocations and disclinations.

The first simulations of small systems of hard discs showed a single first order transition \cite{Alder1962}. Increasing system size a first order transition with short correlation length was ruled out but data were compatible with a weak first-order transition as well as a continuous scenario. Since a phase coexistence was reported in an equilibrated ensemble we would like to argue in favor of a weak first-order scenario. Simulating larger systems with up to 4 million particles with the same computer code it was shown that a Van-der-Waals loop, which is usually interpreted as a first order criterion, weakens with increasing system size \cite{Mak2006}. But especially in small systems the existence of a Van-der-Waals loop can not be taken as solely criterion since the size of the loop strongly depends on the boundary conditions and even systems wich are known to have continuous transitions show a Van-der-Waals loop \cite{Alonso1999}. Simulations of dipolar particles were consistent with KTHNY theory \cite{Lin2006}. Wether a system melts via a first order or via KTHNY theory may depend on the core energy of dislocation \cite{Strandburg1988} a quantity which one can calculate a priory only in the dilute limit of dislocations where renormalization effects do not appear \cite{Sengupta2000}. Binder et al. pointed out that the KTHNY scenario may easily be preempted by a first order transition \cite{Binder2002}. Recent large scale simulations of hard core particles reported a continuous transition between crystal and hexatic phase but since a phase coexistence of hexatic and isotropic liquid was found, this transition is reported to be first order \cite{Bernard2011}.

In experimental systems, the existence of the intermediate hexatic phase is well established \cite{Marcus1997,Kusner1994,Zahn1999,Zahn2000,Segalman2003,Angelescu2005,Keim2007,Han2008} but the nature of the transition is debated as well. Indications of first order transitions are reported in a colloidal system with screened coulomb interactions \cite{Marcus1997} and di-block copolymer systems \cite{Segalman2003,Angelescu2005}. Like in simulations, a phase coexistence is usually interpreted as fist order signature. KTHNY theory is a melting theory starting from large single crystals but of cause a thermodynamic phase should be independent of the history of the matter and cooling and heating cycles should serve the same results. Wang et al. reported to find poly-crystalline domains during cooling in a system of diameter-tunable microgel spheres at finite cooling rates \cite{Wang2010}. Indeed, cooling rates have to be small such that critical fluctuations of continuous order phase transition can switch the symmetry globally. In the present system of particles with dipolar interaction we found (within the given resolution of temperature) both transitions to be continuous during melting and freezing. If the cooling rate is very slow (keeping the system always in quasi-thermal equilibrium) we do observe large single crystal domains in the field of view \cite{Keim2007} implying that KTHNY theory also holds for freezing. But when cooling the system rapidly from the isotropic liquid to the crystalline state we find a poly-crystalline sample without a signature of a hexatic phase during crystallization \cite{Dillmann2008}. But even if growing crystalline domains are found in a liquid environment shortly after a temperature quench one should not interpret them as liquid - crystal coexistence -- simply because the system is far out of equilibrium. In the same sense one should be careful taking poly-crystallinity solely as a signature of first oder transition if the system is cooled at a finite rate.

The manuscript is organized as follows: after a short introduction about long range order of crystals in two dimensions the melting theory developed by Kosterlitz, Thouless, Halperin, Nelson, and Young is summarized in section \ref{secKTHNY}. In section \ref{secsetup} we introduce our experimental setup and how we realise a colloidal monolayer. The following sections \ref{secg6} -- \ref{secmink} 
introduce several quantities to identify different thermodynamic phases in 2D and discuss the results for our colloidal system with dipolar particle interaction. Finally we summarise the advantages and disadvantages of the measures in the conclusion.

\section{Crystals in two dimension and KTHNY}
\label{secKTHNY}
Since the work of Peierls \cite{Peierls1923,Peierls1935} and Mermin \cite{Mermin1966,Mermin1968} it is known that strictly speaking no crystals exist in systems with dimension $D<3$. In general the significance of fluctuations is increased if the dimension of a system is decreased. Crystal lattices with dimension $D<3$ are thermally unstable due to long-wavelength phonon modes. As a consequence long-range translational order does not exist. In case of $D=2$ Mermin showed that the displacement autocorrelation function
\begin{equation}
\langle [ \mathbf{u} ( \mathbf{r})- \mathbf{u}(\mathbf{r}')]^2\rangle \sim \ln| \mathbf{r}-\mathbf{r}'|  \qquad | \mathbf{r}- \mathbf{r}'|\rightarrow \infty
\end{equation}
diverges logarithmically in the crystalline phase. The slow logarithmic divergence in 2D leads to crystals which posses only a quasi long-range translational order. On the other hand the local crystalline orientation is preserved and a long-range bond orientational order exists. The absence of long-range translational order also affects the shape of the structure factor $S(\mathbf{q})$. In 3D the structure factor is characterized by a number of delta functions
\begin{equation}
S(\mathbf{q}) \sim\delta(\mathbf{q} -\mathbf{G})
\end{equation}
at the reciprocal lattice vectors $ \mathbf{G}$ reflecting a not diverging displacement $ \mathbf{u} ( \mathbf{r})$. In 2D the delta functions are replaced by a set of power-law singularities
\begin{equation}
S( \mathbf{q}) \sim | \mathbf{q} -\mathbf{ G}|^{-2+\eta_{ \mathbf{G}}(T)}
\end{equation}
where
\begin{equation}
\eta_{ \mathbf{G}}(T)=\frac{k_BT| \mathbf{G}|^2(3\mu_R+\lambda_R)}{4\pi \mu_R (2\mu_R + \lambda_R)}
\end{equation}
depends on the Lam\'{e} coefficients $\mu_R$ and $\lambda_R$.

As mentioned before, a 2D crystal is characterized by a long-range orientational and a quasi long-range translation order. The KTHNY theory describe the melting of a hexagonal crystal by the appearance of thermally induced topological defects. The phase transition crystal - hexatic takes place at temperature $T_m$ when thermally generated bounded pairs of dislocations which spontaneously appear in the crystal phase dissociate into single dislocations. While dislocations destroy the quasi long-range translational order a quasi long-range orientational order is preserved. The second phase transition hexatic - isotropic liquid at $T_i > T_m$ occurs when dislocations are separated into free disclinations. Disclinations destroy the remaining quasi long-range orientational order so that in the isotropic liquid both the translational and orientational order is short-range.

By identifying the kind of the translational and orientational order at a given temperature $T$ the state of the system can be defined and thus the phase transition points. The nature of the order is marked in a different behaviour of the density density correlation function $g_G$ and bond orientational correlation function $g_6$. The density density correlation function is given by
\begin{equation}
g_{G}(r=|\mathbf{r}_k-\mathbf{r}_l|)=\langle \rho^{\ast}(\mathbf{r}_k)\rho(\mathbf{r}_l)\rangle
\label{eq_gG}
\end{equation}
where $\rho(\mathbf{r}_k)=e^{i\mathbf{G}\mathbf{r}_k}$ is the translational order parameter of particle $k$ located at position $\mathbf{r}_k$.
Practically the translational correlation function is rarely used to identify the melting temperature $T_m$. This is due to the fact, that in a system with Mermin-Wagner fluctuations being present, the reciprocal lattice vector $\mathbf{G}$ is not easily determined due to the power law singularities of the structure factor. Especially in large systems it is numerically difficult to extract $\mathbf{G}$ unambiguously.

The bond orientational correlation function is defined by
\begin{equation}
g_{6}(r=|\mathbf{r}_k-\mathbf{r}_l|)=\langle \psi_6^{\ast}(\mathbf{r}_k)\psi_6(\mathbf{r}_l)\rangle
\end{equation}
with the sixfold bond orientational order parameter
\begin{equation}
\psi_6(\mathbf{r}_k) = \frac{1}{n_l}\sum_{i=1}^{n_l}e^{i6\theta_{kl}}
\end{equation}
where $n_l$ is the number of nearest neighbors of particle $k$ and $\theta_{kl}$ is the angle between the bond of particles $k$ and $l$ and an arbitrary but fixed reference axis. Here the $\langle\rangle$ brackets correspond to an ensemble average. The long-range orientational order in a crystal is expressed in a long-range bond orientational correlation function $\lim_{r\rightarrow\infty}g_6(r)\neq0$ whereas an algebraic decay of the density density correlation function $g_G(r)\sim r^{-\eta_{ \mathbf{G}}(T)}$ reflects the quasi long-range translational order. The hexatic phase is characterized by an algebraic decay of $g_6(r)\sim r^{-\eta_6(T)}$ (quasi long range orientational order) with an exponent
\begin{equation}
\eta_6(T)=\frac{18k_BT}{\pi F_A}
\end{equation}
depending on Frank's constant $F_A$. The short range translational order on the other hand leads to an exponential decay of $g_G(r)\sim e^{-r/\xi_{ \mathbf{G}}(T)}$ with $\xi_{ \mathbf{G}}(T)$ being the translational correlation length. In the liquid regime of the phase diagram the orientational order is short range too, and the correlation function decays as $g_6(r)\sim e^{-r/\xi_6(T)}$ where $\xi_6(T)$ is the orientational correlation length.

In addition to the predictions of the KTHNY theory different criteria have been proposed to identify the phase transitions. Whereas Wang et al. \cite{Wang2010} tested various 2D freezing criteria in poly-crystalline samples of microgel particles we do so in mono-crystalline samples of dipolar particles. We verify the adaptability of the bond order correlation function, the Larson-Grier criterion, the Lindemann parameter, bond-orientational susceptibility, 2D Hansen Verlet rule, 2D L\"{o}wen-Palberg-Simon criterion, shape factors and Minkowski functionals on the melting transitions.

\section{Experimental System}
\label{secsetup}

\begin{figure}
\centering
\begin{minipage}[c]{.45\linewidth}
\hspace{.5cm}
\includegraphics[width=.98\linewidth]{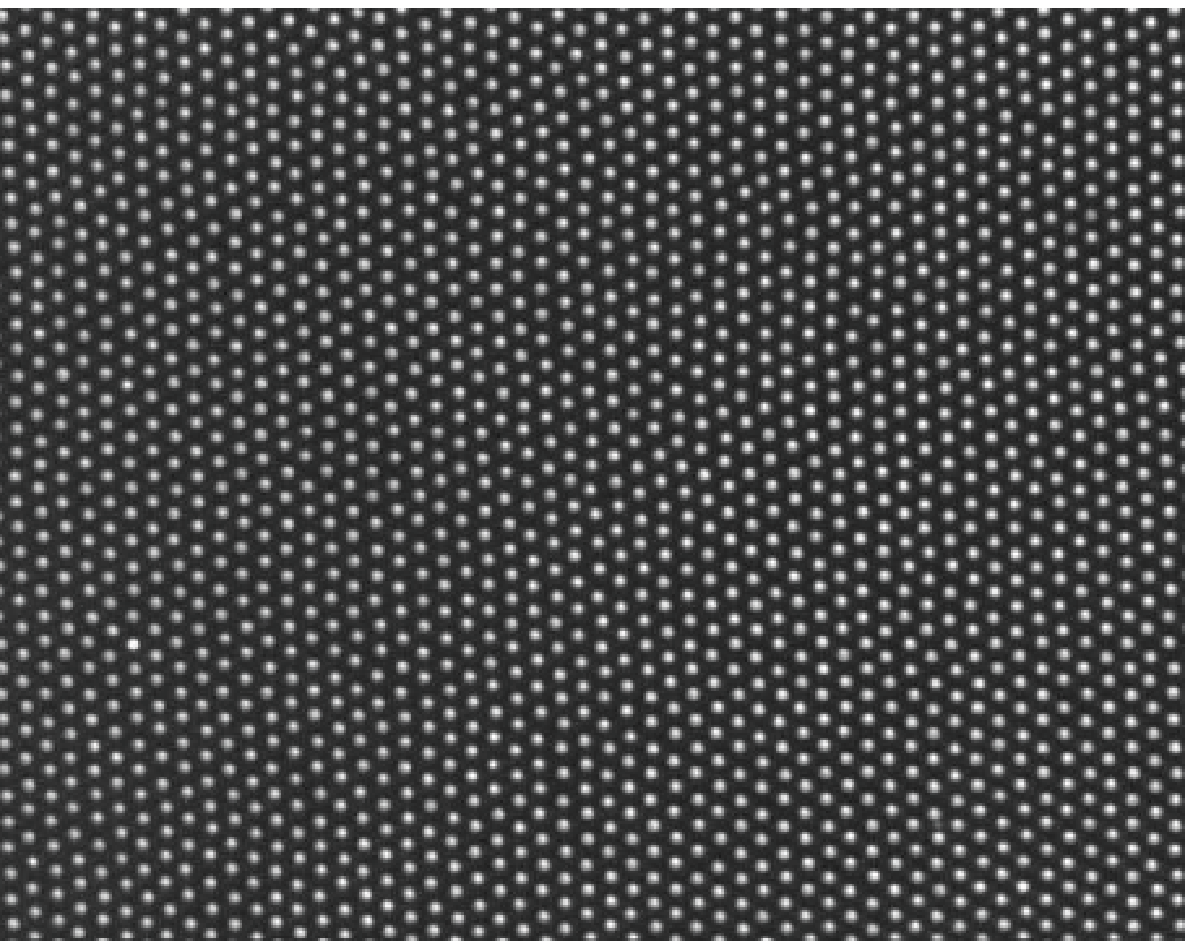}
\end{minipage}
\hspace{.05\linewidth}
\begin{minipage}[c]{.45\linewidth}
\includegraphics[width=1.\linewidth]{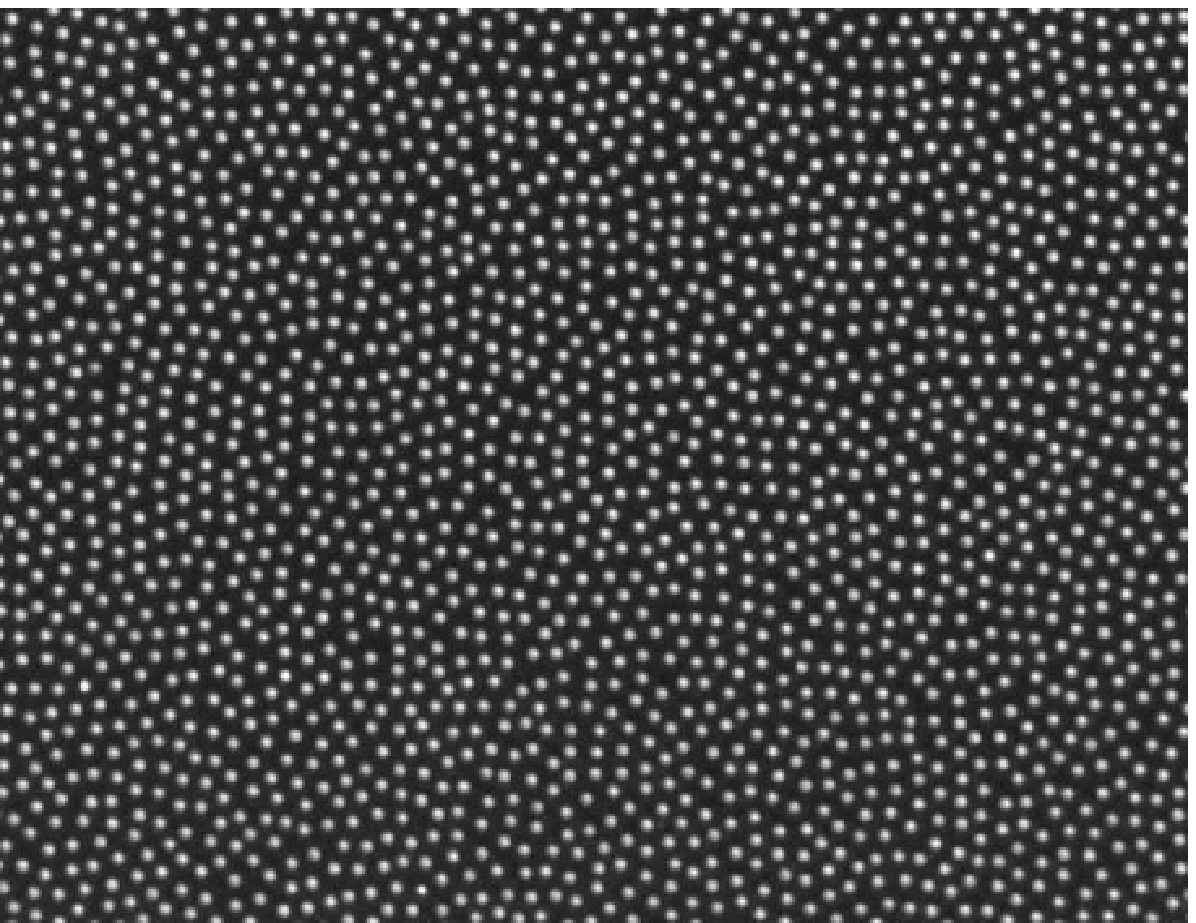}
\end{minipage}
\caption{Cutout of the image of the CCD-camera of a monolayer of colloidal particles. For high magnetic fields particles arrange in a crystal (left) and for low magnetic field particles form a fluid (right). Images are a quarter in size of the full field of view (about $580\times 430~\mu m^2$). The hexatic phase is hardly distinguishable from the isotropic phase solely by eye and ist not shown here.}
\label{xtalandliquid}
\end{figure}

The experimental system is described in detail in \cite{Ebert2009}. It consists of a 2D colloidal monolayer of spherical polystyrene spheres with diameter $d=4.5~\mu m$ suspended in water and sterically stabilized with the surfactant Sodiumdodecylsulfate. Nanoparticles of $Fe_2O_3$ are embedded homogeneously in the polystyrene spheres being responsible for super-paramagnetic behavior and a relatively large mass density of $1.5~g/cm^3$. Therefore particles are confined by gravity at a water/air interface formed by a droplet which is suspended by surface tension in a top sealed cylindrical hole ($6~mm$ diameter) of a glass plate. An external magnetic field $\mathbf{H}$ perpendicular to the water/air interface induces a magnetic moment $\mathbf{M}= \chi \mathbf{H}$ in each particle causing a repulsive dipole-dipole pair-interaction $E_{magn}$ between them. The dimensionless interaction parameter $\Gamma$ which is given by the ratio of the magnetic versus thermal energy
\begin{equation}
\Gamma=\frac{E_{magn}}{k_BT}=\frac{\mu_0}{4\pi}\frac{(\chi \mathbf{H})^2(\pi\rho)^{3/2}}{k_BT}\propto T^{-1}_{sys}
\end{equation}
is equivalent to an inverse system temperature. Under the conditions of temporally constant ambient temperature $T$ and 2D particle density $\rho$ the system temperature depends only on the magnetic field. As a result the system temperature can be easily adjusted by simply changing the strength of the magnetic field $\mathbf{H}$.

An inhomogeneous distribution of the particles within the sample would induce a gradient in system temperature causing a spatial dependence of the phase transition. Therefore it is crucial to align the water/air interface absolutely planar and horizontally. For this purpose several computer controlled regulation loops have been installed to adjust the interface and keep it temporally constant. A monochrome CCD camera is used to observe particles by video microscopy. The field of view ($1158\times 865~\mu m^2$) contains $\approx9000$ particles whereas the whole system includes $\approx250000$ particles. During data acquisition the coordinates of the particles in the field of view are determined in situ every $\approx2~sec$ over a period of $25~min$ by digital image processing with an accuracy of about $50~nm$. This way the phase space information is accessible on all relevant length and time scales.
A crystal is melted by stepwise increasing the system temperature via a reduction of the magnetic field. In the range of the phase transitions the interaction parameter was changed in small steps with an increment of $\Delta\Gamma\approx0.25$. After each modification of the interaction parameter the system was equilibrated for about a day. Figure \ref{xtalandliquid} shows images of the colloidal monolayer in the crystalline (left) and isotropic fluid phase (right).\\

\section{Orientational correlation function}
\label{secg6}

\begin{figure}
\centering
\includegraphics[width=12.cm]{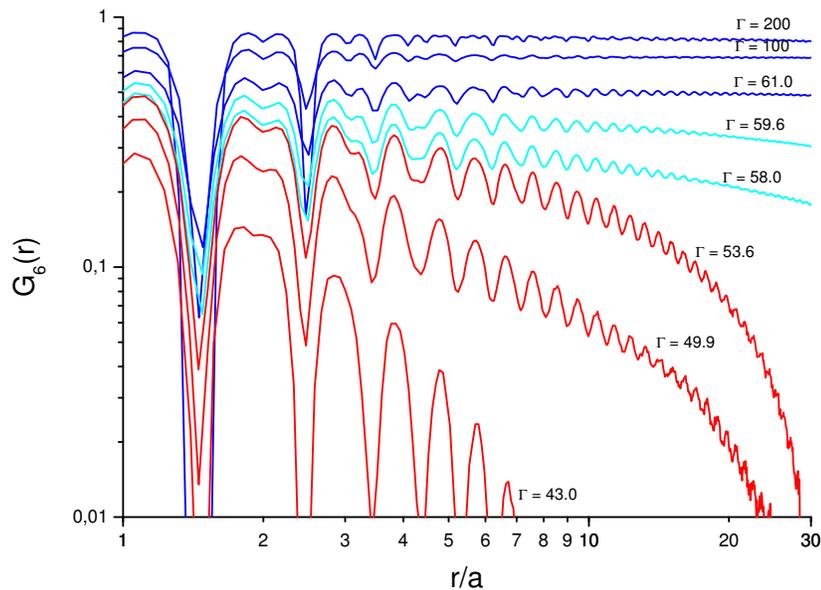}
\caption{Orientational correlation functions $g_6(r)$ (in units of particle distances $a$) for different interaction parameters $\Gamma$. For $\Gamma\geq60$, $g_6(r)$ reflects the long-range order of a crystal (upper three curves) while in the hexatic phase an algebraic decay is observed (two curves in the middle). An exponential decay is observed in the isotropic liquid (lower three curves).}
\label{g6}
\end{figure}
As key-quantity of KTHNY theory, figure \ref{g6} shows the bond orientational correlation function for temperatures in the crystalline, hexatic and isotropic liquid regime. The evaluation of $g_6(r)$ includes a time average over several particle configurations in addition to the ensemble average. The bond correlation function stays finite for interaction parameters $\Gamma>\Gamma_M = 60$ in the crystalline phase. An algebraic decay is observed for $\Gamma=59.6$ and $\Gamma=59.3$ so that the phase transition crystalline - hexatic takes place between $\Gamma=60$ and $\Gamma=59.6$. In the liquide state ($\Gamma=47.6$ and $\Gamma=39.5$) the decay of $g_6(r)$ is exponential. This behavior is in accordance with the predictions of the KTHNY theory and reconfirms previous experimental results \cite{Zahn1999,Zahn2000,Keim2007} for the given system. The hexatic - isotropic transition can be determined by the investigation of the goodness of fit statistics of algebraic or exponential decay \cite{Keim2007}. In principle one could also determine the hexatic - crystalline transition by investigating the slope of $g_6(r)$ in a log-log plot but it is numerically not very precise to distinguish between a small but finite and zero slope decay. From $g_6(r)$ one can also extract two diverging quantities: The orientational correlation length diverges at $\Gamma_i$ and Frank's constant diverges at $\Gamma_m$. In \cite{Keim2007} we fitted both divergencies to extract critical exponents but the transition temperatures were not taken as fitting parameter but used as an input. Otherwise fitting the transition temperatures to a single sided divergence in a finite field of view always overestimates the transition points by a few per cent.

\section{Local bond-order by Larson-Grier}
\label{secLG}

\begin{figure}
\centering
\includegraphics[width=12.cm]{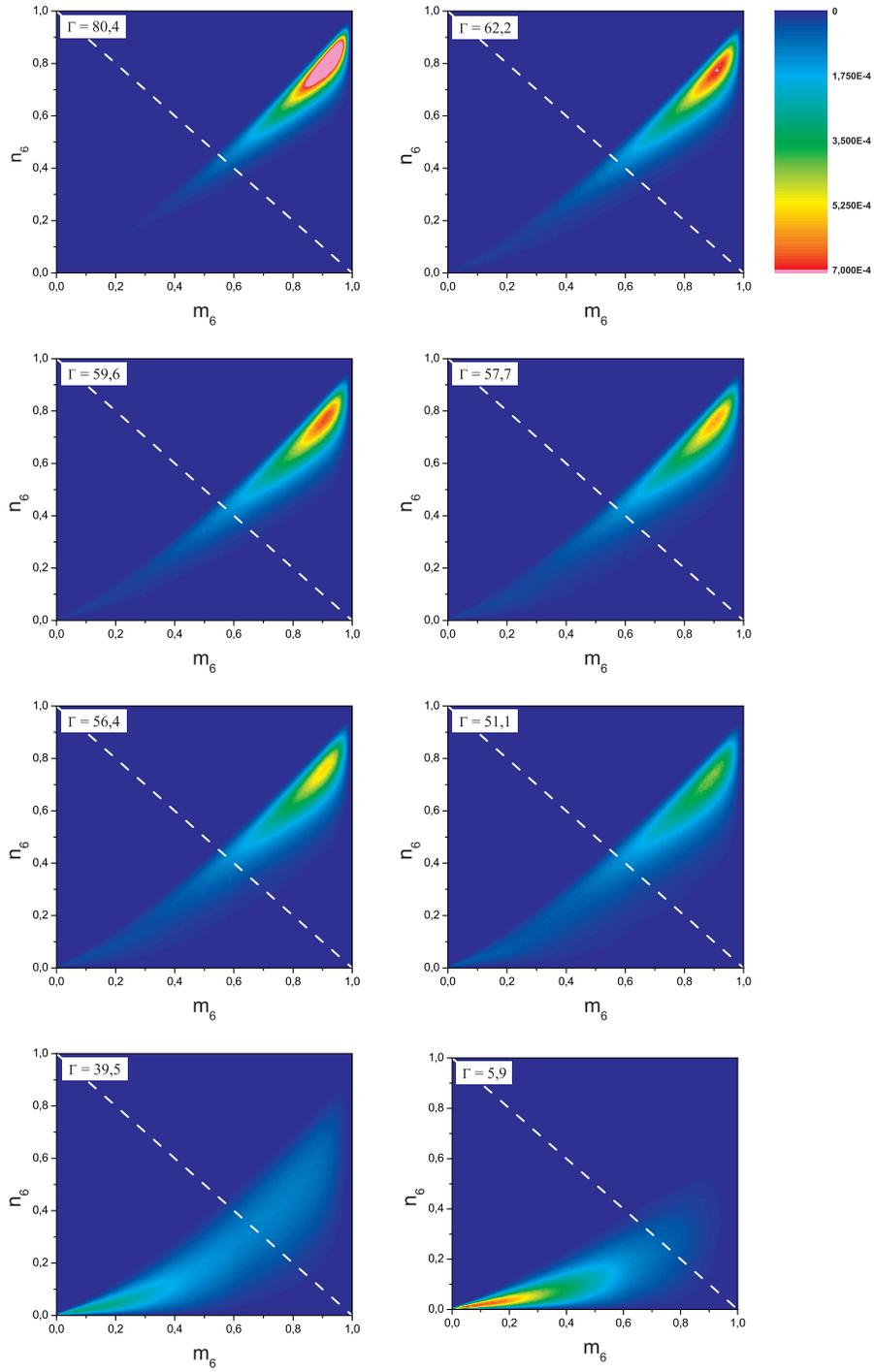}
\caption{Probability distribution of the magnitude of the local bond order parameter $m_6$ versus the magnitude of the projection of $\psi_6$ to the mean of nearest neighbors $n_6$. The probability distribution changes continuously at both phase transitions and no bimodal distribution can be found.}
\label{mnplane}
\end{figure}

To get insight into the local symmetry we focus on the magnitude of the local bond order parameter
\begin{equation}
\label{locpsi} m_{6_k} = |\psi_{6}(\mathbf{r}_k)| \quad .
\end{equation}
$m_{6_k}$ is zero for perfect five- or sevenfold neighbored particles and one for perfect sixfold ones. It measures how the neighbors of particle $k$ fit locally on a hexagonal lattice. To compare the local sixfold symmetry with neighboring particles Larsen and Grier \cite{Larsen1996} investigated the magnitude of the projection of $\psi_{6_k}$
\begin{equation}
\label{locpsi} n_{6_k} = |\psi_{6_k}^\ast* 1/N_l\sum_l\psi_{6_l}|
\end{equation}
to the mean local orientation field. It takes the second nearest neighbors into account and determines how the orientation of
particle $k$ fits into the orientation of its neighbor particles. Since it is a projection $n_{6_k} \leq m_{6_k}$ and $n_{6_k} +
m_{6_k} \leq 2$. In \cite{Larsen1996} an uni-modal distribution was found even if real space images showed a dilute liquid (or gas) phase and dense crystalline flakes implying an attractive interaction between particles to exist, whereas in \cite{Marcus1997} a bimodal distribution is reported next to the isotropic-hexatic as well to the hexatic-crystalline transition. Particles in the $m_6$-$n_6$-plane with $m_6+n_6>1$ (upper right corner) where identified to be crystal-like particles. Figure \ref{mnplane} shows the probability distribution for our purely repulsive system in the $m_6$-$n_6$-plane for several temperatures.  The upper row are plots in the crystalline phase, the second row shows plots of the hexatic phase whereas the two lowest rows are all from fluid phases. The absence of a bimodal distribution and the weak dependence of the local bond-order field above and below $\Gamma_i$ and $\Gamma_m$ indicate continuous phase transitions. The third row shows that the local order in a 2D fluid is predominately hexagonal, even far away from the phase transitions. Only at very high temperatures (low interactions strength) most of the particles have $m_6+n_6<1$ indicating that the local six-fold order up to the second shell is lost (lowest line). Since the dependence of the local bond order on the different phases is weak, it does not serve as a sharp criterion for phase transition temperatures.

\section{Lindemann parameter}
\label{secLinde}

The Lindemann parameter is a well-known criterion in 3D to identify the melting point of crystalline structures. According to Lindemann \cite{Lindemann1910} the melting of a crystal takes place if the thermal energy leads to displacements of atoms in relation to their equilibrium lattice sites which are in the range of one-half of the interatomic distance. The Lindemann criterion was modified by Gilvarry \cite{Gilvarry1956} by considering the root-mean-square amplitude of thermal vibrations. He suggests that the melting process is initiated when the fraction of the root-mean-square amplitude and the interatomic distance reaches a critical value of approximately $0.1$.

\begin{figure}
\centering
\includegraphics[width=12.cm]{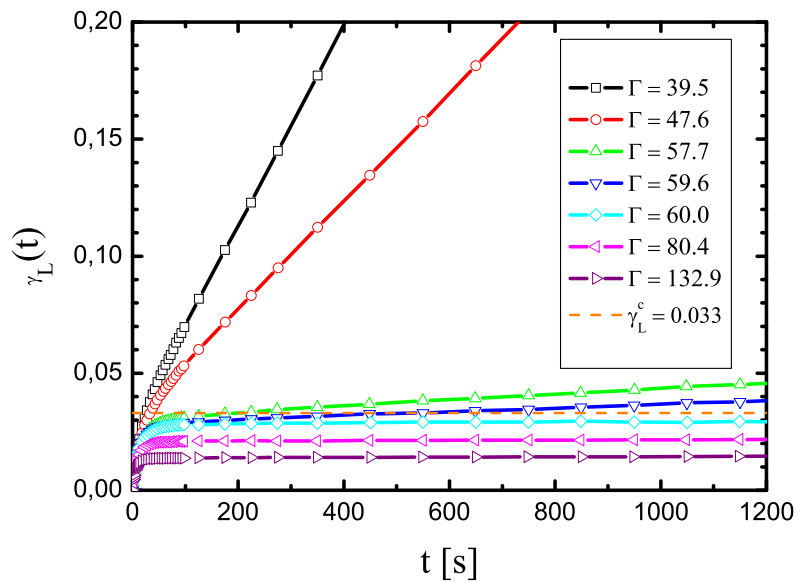}
\caption{The dynamic Lindemann parameter stays finite in the crystal phase and diverges in the hexatic and isotropic liquid state. $\Gamma$ increases from top to bottom with the same order as in the label.}
\label{lindemann}
\end{figure}

The Lindemann criterion in this form is inapplicable in 2D. Due to long wavelength phonon modes the Lindemann parameter diverges in a crystal as well as in a liquid. Bedanov et al. \cite{Bedanov1985} introduced a melting criterion for 2D
\begin{equation}
\gamma_m=\frac{\langle |\mathbf{u}_j-\mathbf{u}_{j+1}|^2 \rangle}{a^2} \label{equationlinde}
\end{equation}
analog to the Lindemann parameter in 3D based on the displacement $\mathbf{u}_j$ of particle $j$ with respect to its nearest neighbors $j+1$ and normalised to the average inter-particle distance $a$. Zahn et al. \cite{Zahn1999} generalised equation (\ref{equationlinde}) to a dynamic Lindemann parameter
\begin{equation}
\gamma_L(t)=\frac{\langle[\Delta\mathbf{u}_i(t)-\Delta\mathbf{u}_{i+1}(t)]^2\rangle}{2a^2}
\end{equation}
where $\Delta\mathbf{u}(t)= \mathbf{u}(t)-\mathbf{u}(t=0)$. The crystal - hexatic phase transition can be determined by the long time behaviour of $\gamma_L(t)$. The Lindemann parameter $\gamma_L(t\rightarrow\infty)$ diverges in the hexatic and isotropic liquid phase whereas in a crystal $\gamma_L(t\rightarrow\infty)$ stays finite below a critical value $\gamma_L^c=0.033$.

The Lindemann parameter $\gamma_L(t)$ is shown in figure (\ref{lindemann}) for different interaction parameters $\Gamma$. As expected $\gamma_L(t)$ converges in a crystal ($\Gamma\geq60$) to a finite value $<\gamma_L^c$ but diverges in the hexatic phase ($\Gamma=59.6$ and $\Gamma=57.7$) and isotropic liquid state ($\Gamma=53.7$, $\Gamma=49.6$ and $\Gamma=39.5$). According to the behaviour of the Lindemann parameter the transition crystal - hexatic occurs in the range between $\Gamma=60.0$ and $\Gamma=59.6$. This result is in excellent agreement with the predictions of the KTHNY theory and the determination of the phase transition crystal - hexatic obtained with the help of the bond orientational correlation function $g_6$. If grain boundaries are visible due to finite cooling rates or density gradients in the sample, the dynamic Lindemann parameter is not finite for the crystalline state. But for a mono-crystalline sample it acts as a very sensitive tool to determine the crystal - hexatic phase transition temperature $\Gamma_m$.

\section{Bond-orientational susceptibility}
\label{secsusc}

\begin{figure}[b]
\centering
\includegraphics[width=12.cm]{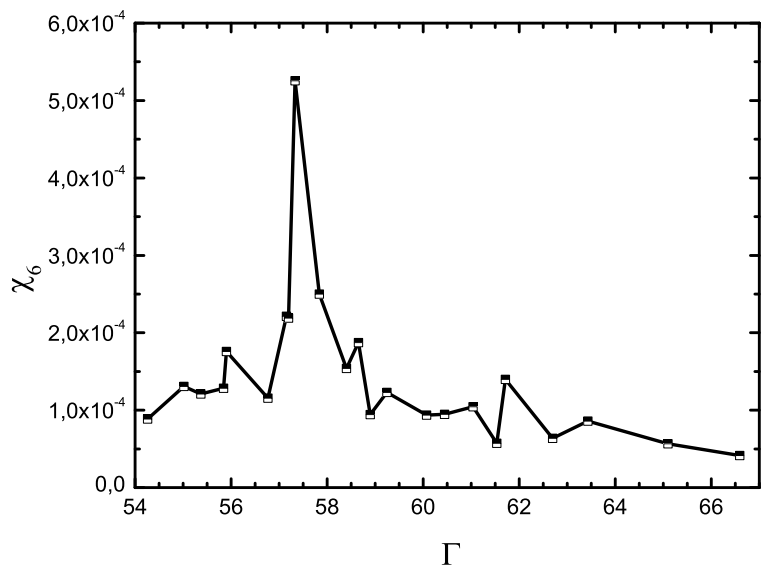}
\caption{The bond-orientational susceptibility for interaction parameters in the isotropic liquid, hexatic and solid state. The maximum of the peak corresponds to the isotropic liquid phase transition.}
\label{chi6}
\end{figure}

The hexatic - isotropic liquid phase transition is associated with fluctuations of the orientational order parameter $\psi_6$. The fluctuations can be quantified by the bond-orientational susceptibility
\begin{equation}
\chi_6 = A(\langle|\Psi_6^2|\rangle - \langle|\Psi_6|\rangle^2)
\end{equation}
where $\Psi_6=1/N\sum_{k=1}^N \psi_6(\mathbf{r_k})$ is the global bond orientational order parameter of the $N$ particles included in a system with size $A$. The bond-orientational susceptibility increases dramatically if the temperature reaches the point of the hexatic - isotrop liquid phase transition at $\Gamma_i$ \cite{Weber1995}. Here an increase of $\chi_6$ is observed independently whether the system approaches the transition point from the liquid $\Gamma\rightarrow\Gamma_i^-$ or from the hexatic phase $\Gamma\rightarrow\Gamma_i^+$. This behavior of the bond-orientational susceptibility simplifies the identification of the hexatic isotropic liquid phase transition in comparison with the previously mentioned method of the single sided divergence of the orientational correlation length $\Gamma\rightarrow\Gamma_i^-$. In figure \ref{chi6} we see a sharp increase of $\chi_6$ at $\Gamma=57.5\pm0.5$. This result coincides with the value for the hexatic isotropic transition obtained in \cite{Keim2007}. The bond-orientational susceptibility is a very sensitive tool to determine the transition temperature since we find a sharp increase from both sides of the peak (unlike e.g. the divergence of the orientational correlation length calculated from $g_6(\mathbf{r})$ where the divergence is single sided, see section \ref{secg6} or \cite{Keim2007}). In principle one could use the bond-orientational susceptibility as criterion for first order or second order transition \cite{Weber1995}. A symmetric peak shape is predicted for second order or continuous transition whereas for first order transitions the limit of the susceptibility from above and below the transition should be different. Due to the limited temperature resolution in our data we do not want to overestimate this topic but on a first glance the data seem to be consistent with a second order transition.

\section{Structure factor}
\label{secSq}

\begin{figure}
\centering
\includegraphics[width=12.cm]{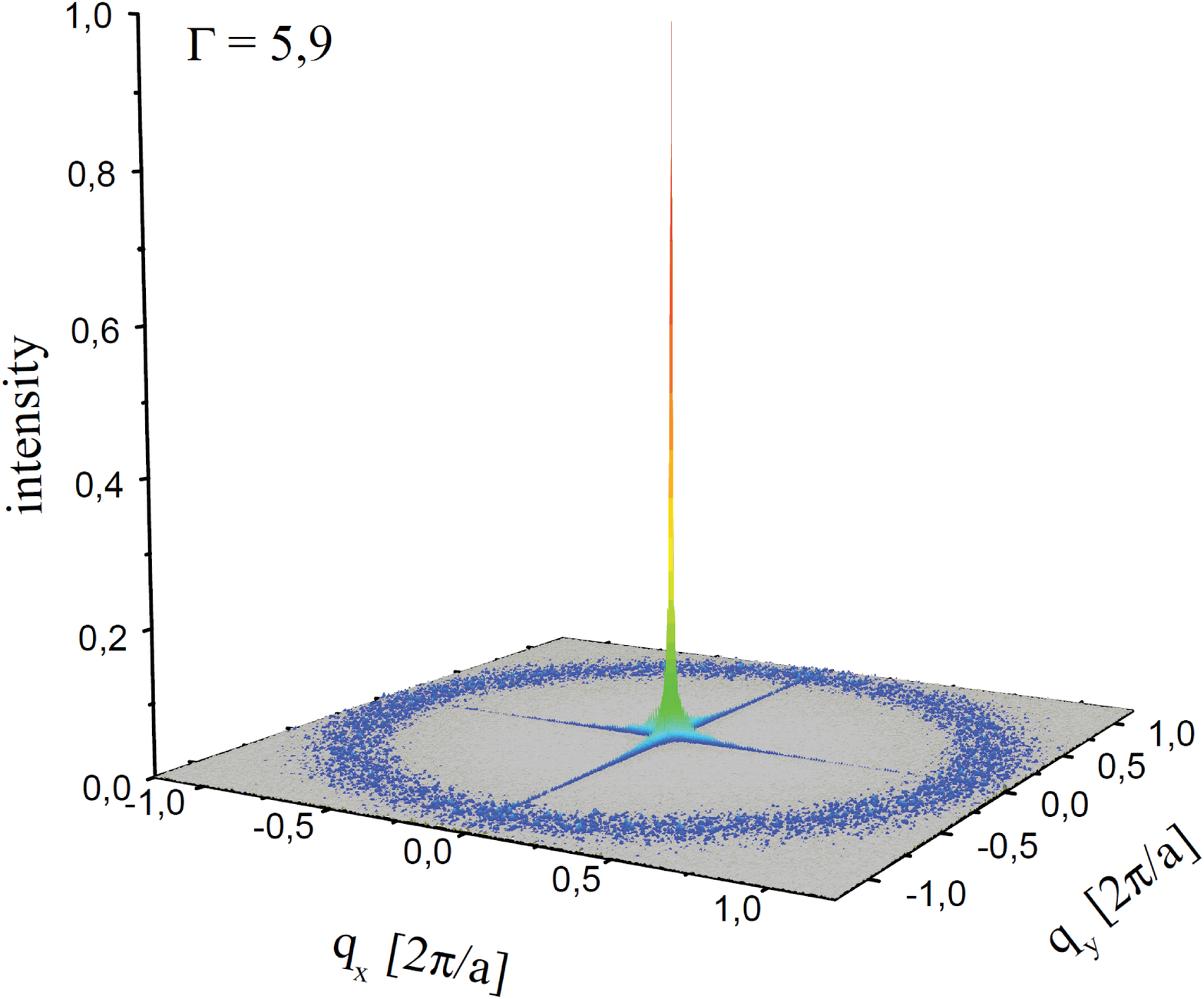}
\vspace{1cm}

\includegraphics[width=12.cm]{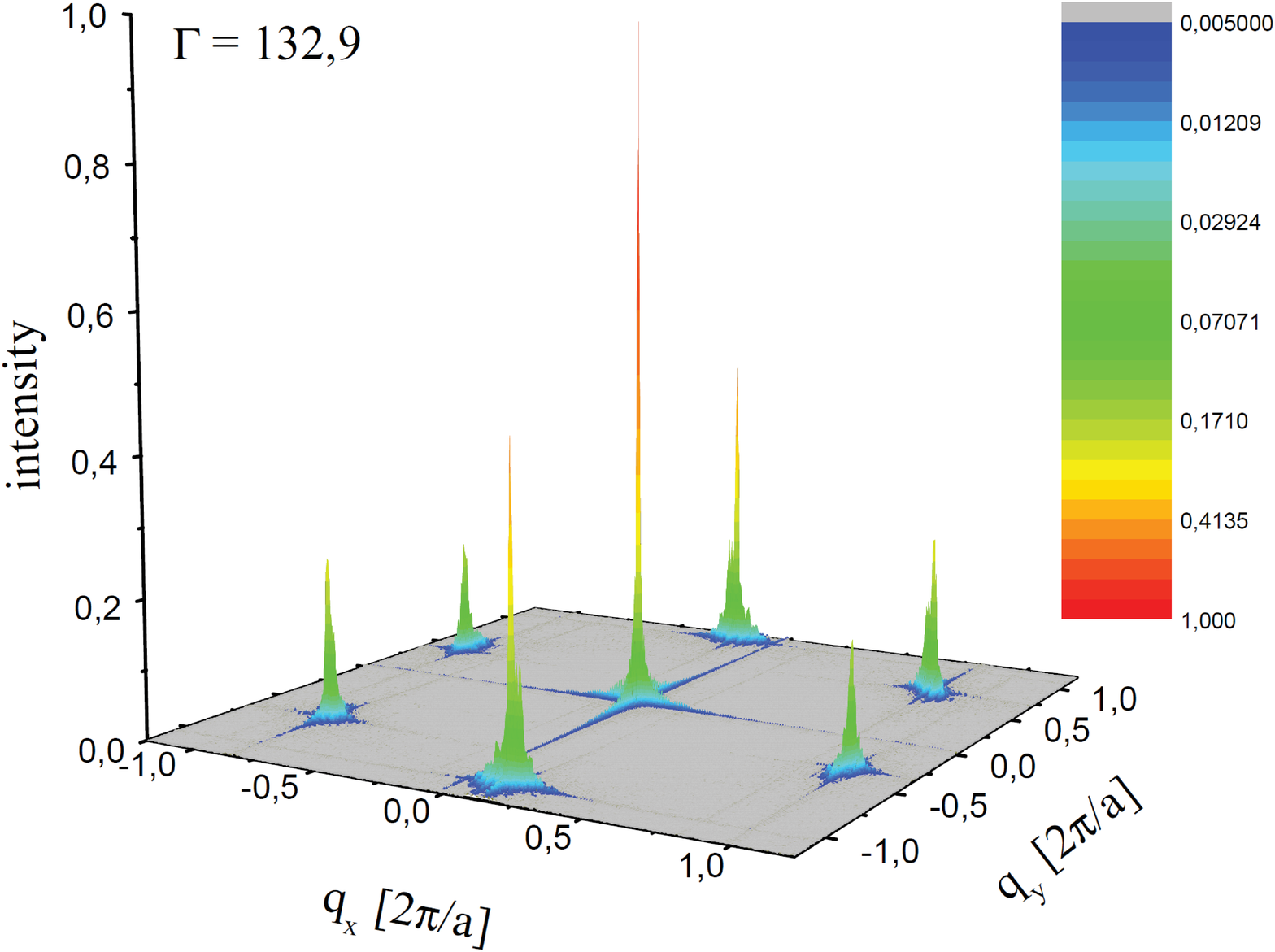}
\caption{The structure factor $S(\mathbf{q})$ in the liquid phase (top) and crystalline phase (bottom). The rectangular cross in the center is an artifact due to the finite field of view.}
\label{sqvec}
\end{figure}

The structure factor is another often used physical quantity to identify the freezing transitions. It is defined by
\begin{equation}
S(\mathbf{q}) = \frac{1}{N}\langle\rho(\mathbf{q})\rho(-\mathbf{q})\rangle
\end{equation}
where the spatial Fourier transform  of the number density $\rho(\mathbf{q})$ is given by
\begin{equation}
\rho(\mathbf{q})= \sum_{i=1}^N \exp(i\mathbf{q}\mathbf{r}_i) \quad .
\end{equation}
Here $\langle\rangle$ denotes an ensemble average over $N$ particles located at positions $\mathbf{r}_i$. Since we know the time dependent trajectories of the particles in the field of view we calculate the structure factor as function of time
\begin{equation}
S(\mathbf{q},t) = \frac{1}{N}\sum_{i=1}^N\sum_{j=1}^N \exp(i\mathbf{q}\mathbf{r}_i(t)) \exp(-i\mathbf{q}\mathbf{r}_j(t))
\end{equation}
and determine $S(\mathbf{q})=n_t^{-1}\sum_{t=1}^{n_t} S(\mathbf{q},t)$ by a time average over $n_t>70$ statistically independent particle configurations. \\

\begin{figure}
\centering
\includegraphics[width=12.cm]{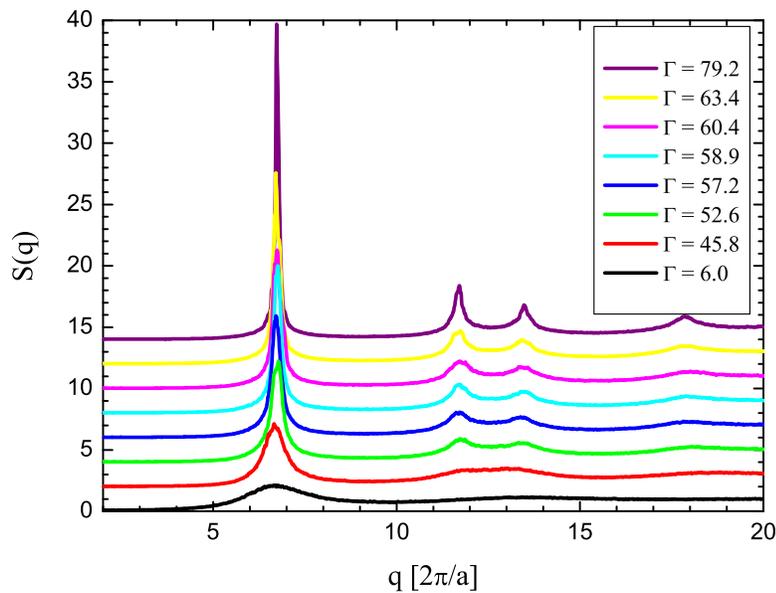}
\caption{The isotropic structure factor $S(q)$ for different interaction parameters. Curves are shifted for reasons of clarity, $\Gamma$ decreases from top to bottom with the same order as in the label.}
\label{sq}
\end{figure}

\begin{figure}
\centering
\includegraphics[width=12.cm]{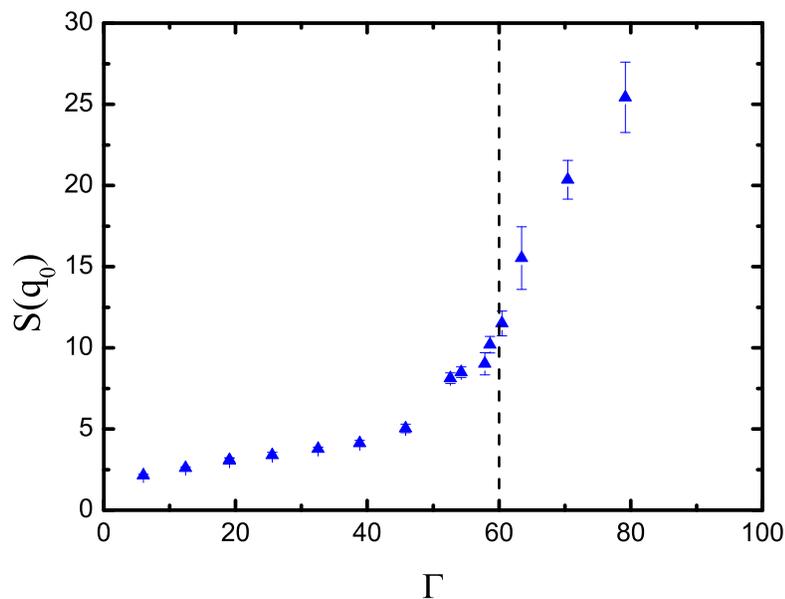}
\caption{Amplitude of the first maximum of the structure factor $S(q_0)$. At the freezing point $\Gamma=60$ the amplitude $S(q_0)\simeq10$ for the system with long range particle interaction. Two error-bars are shown in the isotropic and crystalline phase, calculated as time average from different time steps at the given temperature.}
\label{sqmax}
\end{figure}

Figure \ref{sqvec} show the structure factor $S(\mathbf{q})$ in the isotropic liquid and the crystalline phase calculated from particle trajectories. Performing an azimuthal average gives the classical structure factor $S(q)$. Hansen and Verlet argued that freezing is associated with the amplitude $S(q_0)$ of the first maximum of the isotropic structure factor. A $3D$ liquid freezes when $S(q_0)$ exceeds a characteristic value of $2.85$ \cite{Hansen1969}. The predictions of the characteristic value in $2D$ resulting from simulations vary from $S(q_0)=4.4$ for particles with hard core and coulomb interaction \cite{Caillol1982} to $S(q_0)=5.75$ \cite{Ramakrishnan1982} for particles with $r^{-12}$-interaction. Figure (\ref{sq}) shows the temperature dependent isotropic structure factor $S(q)$ which is obtained by an angular average of the structure factor $S(\mathbf{q})$. The maximum of the isotropic structure factor $S(q_0)$ increases continuously as the temperature decrease (increasing interaction parameter $\Gamma$). Additionally the second maximum splits in two peaks which reflects an evolving hexagonal structure. As shown in figure (\ref{sqmax}) $S(q_0)$ rises slowly in the liquid and hexatic phase followed by a sharp rise after crossing the freezing point $\Gamma=60$. The characteristic value is $S(q_0)\simeq10$ and thus exceeded the estimated value by a factor of $\approx2$ for our system with long range dipole-dipole interaction between the particles.

\begin{figure}
\centering
\includegraphics[width=12.cm]{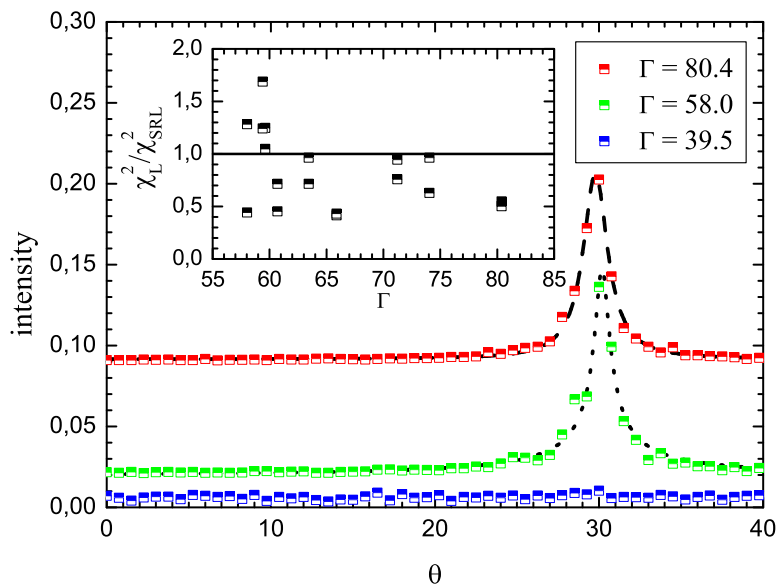}
\caption{The line shape of a Bragg peak in the solid ($\Gamma=80.4$) and the fit with a Lorentzian function (dashed line) and in the hexatic phase ($\Gamma=58.0$) with a SQL function fit (dotted line). In the isotropic liquid ($\Gamma=39.5$) the intensity shows no angular dependence. The curves are shifted for clarity. The inset shows the ratio of the goodness-of-fit-statistic as function of system-temperature. Above $\Gamma_m = 60$ Loerentzian function fits better, below $\Gamma_m$ a SQL function fits better (except the data-point at $\Gamma = 58$, see main text).}
\label{lineshape}
\end{figure}

Another criterion about the global order is given by the line shape of the angular intensity of the structure factor. According to \cite{Aeppli1984,Davey1984} the line shape of the Bragg peaks in the solid state are given by a Lorentzian function $S(\theta_0)=[(\theta_0-\theta)^2+\kappa^2]^{-1}$ where $\theta_0$ is the angular position of the maximum of a Bragg peak, $\kappa$ is the angular width of the Lorentzian function, and the in-plane angle $\theta$ ranges from $\theta-\pi/6$ to $\theta+\pi/6$ because of the sixfold symmetry. In the hexatic phase a square-root Lorentzian (SRL) behavior $S(\theta_0)=[(\theta_0-\theta)^2+\kappa^2]^{-1/2}$ is expected. The line shapes of a Bragg peak in the solid $\Gamma=80.4$ and hexatic $\Gamma=58.0$ phase are shown in figure \ref{lineshape} whereas in the isotropic liquid state $\Gamma=39.5$ no angular dependence of the intensity is observed. To evaluate the behaviour of the line shapes for different interaction parameters in the solid and hexatic phase a fit with a Lorentzian as well as a square-root Lorentzian function was executed. The line shape was determined on the basis of the reduced chi-square goodness-of-fit statistic $\chi^2$ of the fits.
The ratio $\chi^2_L/\chi^2_{SRL}$ of the Lorentzian and square-root Lorentzian reduced chi-square goodness-of-fit statistic is given in the inset of figure \ref{lineshape}.
The line shapes in the solid state are well reproduced by a fit with a Lorentzian function $\chi^2_L/\chi^2_{SRL}<1$ while in the hexatic phase a square-root Lorentzian function fits better $\chi^2_L/\chi^2_{SRL}>1$. Only in the vicinity of the hexatic to isotropic liquid phase transition at $\Gamma_i= 57.5$ this is not the case for the data-point at $\Gamma = 58$. This might be due to the fact that below $\Gamma_i$ the peaks should disappear and this datapoint is to close to the isotropic transition to distinguish unambiguously between Lorentzian and square-root Lorentzian azimuthal shape.

\section{L\"{o}wen-Palberg-Simon criterion}
\label{secLPS}

\begin{figure}[b]
\centering
\includegraphics[width=12.cm]{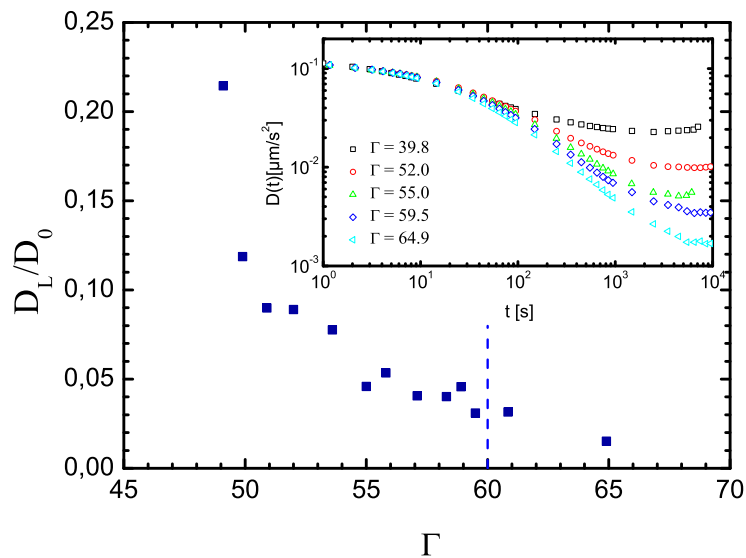}
\caption{The temperature dependent ratios of the short-time versus long-time self-diffusion coefficients $D_L/D_0$. At the freezing point $D_L/D_0\approx0.03$ which has to be compared to a value of $D_L/D_0=0.086$ expected from simulation. The inset shows the time dependent diffusion constant calculated for different time-windows from the mean squared displacement.}
\label{lps}
\end{figure}

L\"{o}wen, Palberg and Simon \cite{Lowen1993} introduced a freezing criterion based on the dynamical properties of $3D$ systems. Their criterion states that a system starts to solidify if the ratio of the long-time self-diffusion coefficient $D_L$ and the short-time self-diffusion coefficient $D_0$ reaches a critical value of $0.1$. Brownian dynamics simulations of different pair potentials lead to critical values in $2D$ between $0.072$ (hard disks) and $0.099$ ($r^{-12}$ potential). In case of a dipolar interaction a critical value of $0.086$ was obtained \cite{Lowen1996}.
The short-time and long-time self-diffusion coefficients are related to the mean-square displacement $\Delta r^2 = 1/N\sum_{i=1}^N[r_i(t)-r_i(0)]^2$ by the following equations:
\begin{eqnarray}
D_0=\lim_{t\rightarrow 0} \frac{\Delta r^2}{4t} \\
D_L=\lim_{t\rightarrow \infty} \frac{\Delta r^2}{4t}
\end{eqnarray}
The ratio of the short-time and long-time self-diffusion coefficients for different interaction parameters are shown in figure \ref{lps}. The critical value expected for the freezing point from simulation with a dipolar interaction can not be reproduced. At the solid hexatic phase transition $D_L/D_0\approx0.03$ which is a factor of three smaller compared to computer simulation for about 1000 dipolar particles \cite{Lowen1996}. In a poly-crystalline system of soft spheres a threshold of about $D_L/D_0 = 0.08$ is reported \cite{Wang2010} supporting the simulations. Since we know that grain boundaries influence the 2D dynamical Lindemann parameter we expect a difference in the long time diffusion coefficient between systems with and without grain boundaries. Additionally the short time diffusion coefficient of particles at a water/air interface is larger compared to particles in the bulk. This together with the fact that grain boundaries are not visible in our field of view might explain the small value of $0.03$ in our system. Zippelius \cite{Zippelius1980} pointed out, that the dependence of the ratio $D_L/D_0$ as function of temperature may be used as criterion for first and second order transition. First oder transitions are characterized by a discontinuous jump at $T_m$ whereas a smooth change should be found for KTHNY-like behaviour. Figure \ref{lps} shows a continuous variation as function system temperature. This again supports the result that our system melts according to KTHNY theory.

\section{Shape factor}
\label{secshape}
\begin{figure}[t]
\centering
\includegraphics[width=12.cm]{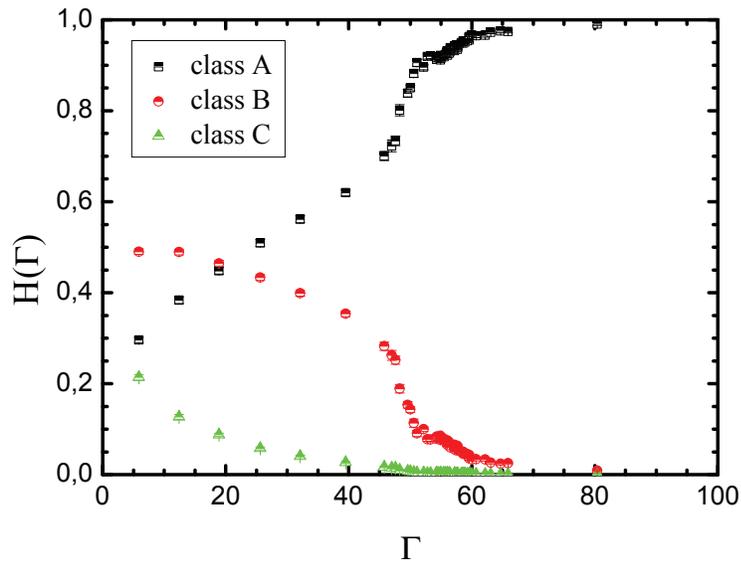}
\caption{The fraction of particles assigning to three different classes $A$,$B$ and $C$. Error-bars (calculated as time average for different time steps at a given temperature) are smaller than the symbols.}
\label{class}
\end{figure}

\begin{figure}[b]
\centering
\includegraphics[width=12.cm]{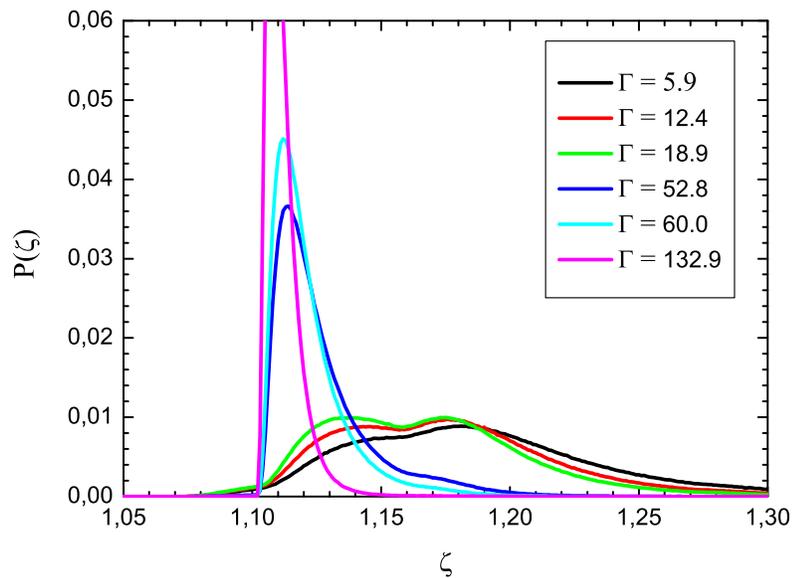}
\caption{The distribution of shape factors $P(\zeta)$ for different interaction parameters. The distribution changes from a bimodal shape for $\Gamma<60.0$ to a unimodal shape in the solid $\Gamma\geq60$.}
\label{shape}
\end{figure}

Mou\v{c}ka and Nezbeda introduced a shape factor $\zeta$ to analyze structural changes occurring for simulations of a hard disk system in the region of phase transitions \cite{Mouifmmodecheckcelsevcfika2005}. The shape factor of particle $i$ is defined as
\begin{equation}
\zeta_i=\frac{C_i^2}{4\pi S_i}
\end{equation}
where $C_i$ and $S_i$ corresponds to the perimeter and area of the Voronoi cell of the particle. The shape factor for a regular polygon with $n$ edges is given by $\zeta_n^{reg}=n/\pi \tan(\pi/n)$. They observed that the distribution of the shape factor $P(\zeta)$ becomes bimodal near the freezing point. The distribution in the liquid state is broad and the maximum is located at relative high $\zeta$-values whereas in the solid a sharp distribution with a maximum near the value of a regular hexagon $\zeta=1.03$ is given. This behaviour is explained by the evolution from different type of cells which are also distorted in the liquid to more regular hexagonal cells in the solid. Reis et al. \cite{Reis2006} classified the particles of a granular fluid in three classes according to their shape factor. Particles with $\zeta<\zeta_{min}$ belong to class $A$, particles with $\zeta_{min}<\zeta<\zeta_u$ belong to class $B$ and particles with $\zeta>\zeta_u$ belong to class $C$ where $\zeta_{min}=1.159$ and $\zeta_u=1.25$. If the number of particles in class $A$ and $B$ are equal they observed the transition from a liquid to an intermediate phase (coexistence or hexatic phase could not be evaluated). Approaching the freezing point a sharp decline of the number of class $B$ particles occurs in the intermediate phase. After crossing the freezing point the number of class $B$ particles decreases much more moderately in comparison with the intermediate phase.

Using the classification procedure described in \cite{Reis2006} we receive an identical value for $\zeta_{min}$ and a slightly deviating value $\zeta_u=1.22$. In accordance to the granular system the slope of the fraction of particle class $B$ changes at the freezing point as shown in figure \ref{class}. On the other hand the fractions of classes $A$ and $B$ are equal at $\Gamma\approx20$ far away from the isotropic liquid - hexatic phase transition. We ascribe this different behavior to the structural changes which are less pronounced at the phase transitions compared to the hard disk or granular systems. This can been seen in figure \ref{shape} where even deep in the liquid state at $\Gamma=5.9$ a bimodal distribution $P(\zeta)$ instead of flat one exists. This indicates that in the liquid as well as in the crystalline state a hexagonal configuration of the particles is preferred. This assumption is confirmed by the fact that more than $50\%$ of the particles at $\Gamma=5.9$ are still sixfold coordinated. For our thermal system with long range interaction, the phases are not identified unambiguously using shape factors.

\section{Minkowski functionals}
\label{secmink}

In addition to the previously described criteria we tested tentatively if any hint for phase transitions can be derived from the behaviour of Minkowski measures. Minkowski measures have been successfully adopted to specify spatial patterns, e.g. during spinodal decomposition \cite{Mecke1997}, the evolution of galaxy clusters \cite{C.Beisbart2001} or partial clustering in a $2D$ colloidal glass former \cite{Ebert2009a}. Integral geometry offers a set of topological and geometrical descriptors (Minkowski functionals) to characterize spatial patterns. The operation of Minkowski functional $V$ on patterns $P,Q,..$ have to obey three properties to be a morphological measure:
\begin{enumerate}
  \item Motion invariance: $V(gP+t) = V(P)$ for $g$ and $t$ being any rotation and translation.
  \item Additivity: $V(P\cup Q) = V(P)+V(Q)-V(P\cap Q)$
  \item Continuity: A slightly distortion of a pattern leads to a continuous change of $V$.
\end{enumerate}
According to Hadwiger \cite{Hadwiger1957} exist exactly $D+1$ linear independent Minkowski functionals in $D$ dimension. In case of $D=2$ the Minkowski measures are related to the surface area $A$, the circumference $U$ and the Euler characteristic $\chi=N_a-H$ which is given by the difference of the number of connected surfaces $N_a$ and number of holes $H$.
\begin{figure}
\centering
\includegraphics[width=14.cm]{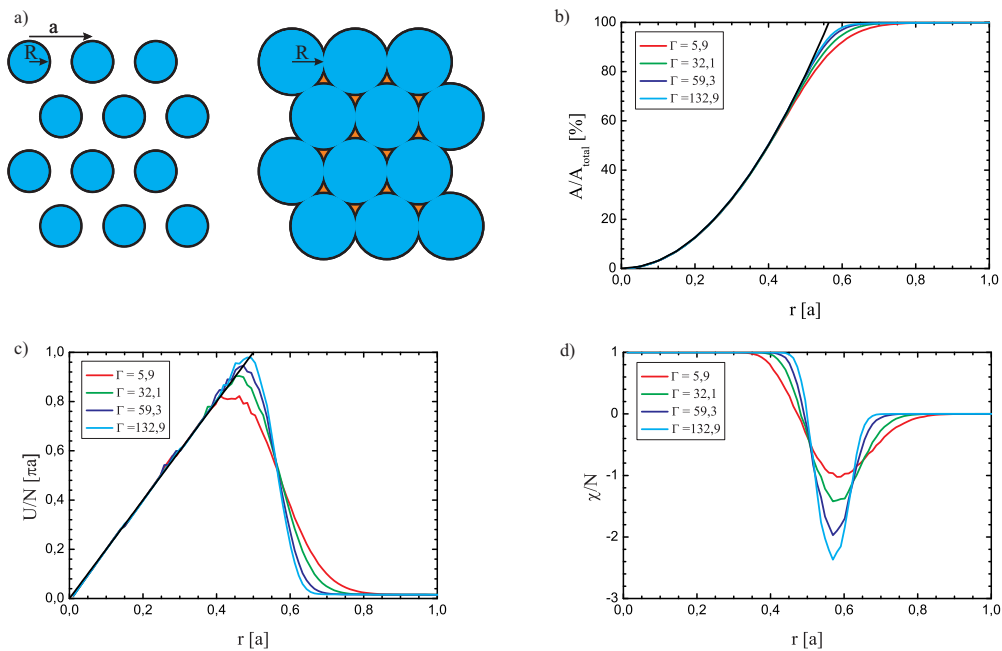}
\caption{In 2D systems three independent Minkowski functionals exist which may be chosen as surface area $A$, the circumference $U$ and the Euler characteristic $\chi=N_a-H$.}
\label{mink}
\end{figure}

We create a pattern for a single particle configuration by placing a cover disk with constant radius $R$ at each particle position (see figure \ref{mink}\;a). Morphological information about the particle configuration is then obtained by determining the Minkowski functionals as a function of cover radius $R$. As crystals in $2D$ posses sixfold symmetry we briefly describe the behaviour of the Minkowski functionals for a perfect hexagonal lattice. The cover disks do not overlap for radius $0\leq R<a/2$. The surface area and circumference are given by the area $A_d(R)$ respectively circumference $U_d(R)$ of a cover disks times the number of cover disks $N$ and the Euler characteristic is $\chi=N$. As requested by additivity for $a/2\leq R<\sqrt{3}a/3$ the surface area is given by sum of the area of the cover disks minus the overlapping areas which is connected with a slower increase of the surface area. Despite the radius is increased the circumference starts to decrease because the parts of the perimeters belong to the overlapping areas of the disks are disregarded. The Euler characteristics gets negative since the overlapping leads to only one connected surface while holes are formed. If the disk radius reaches a value of $R\geq\sqrt{3}a/3$ the whole area is covered and the circumference is equal zero. The holes disappear, i.e. $\chi=1$.

In figure \ref{mink}\;b-d) the three Minkowski functionals $b)$ surface area $A$, $c)$ circumference $U$, and $d)$ Euler characteristics $\chi$ are shown for crystalline, hexatic and isotropic liquid systems. Note that $A/A_{total}$, $U/N$ and $\chi/N$ is plotted. All Minkowski functionals reflect in principle the behaviour which is characteristic for the hexagonal lattice. The deviations from the expected curves increase with the inverse system temperature i.e. with thermal motion. The increase of the surface area is proportional to $R^2$ while the circumference is proportional to $R$ and $\chi=1$ as expected for small radii. The range where holes in the perfect hexagonal lattice exists are broadened to $0.35a\leq R\leq 0.8a$. For $R>0.8a$ the full area is covered, $U/N=0$ and $\chi/N\rightarrow0$. No qualitative changes of the Minkowski functionals can be observed at the phase transitions. Changes of the Minkowski functionals which may emerge due to the emergence of defects can not be identified as their number density is quite small and due to strong thermal fluctuations of the particles. In contrast to a binary dipole-dipole system where structural heterogeneities have been quantified \cite{Ebert2009a} Minkowski functionals are not applicable as order parameter to identify phase transitions temperatures for our mono-disperse system.

\section{Conclusion}
\label{secconcl}

In a two-dimensional system of colloids with repulsive dipolar interaction several criteria based on structural as well as dynamical quantities were compared to identify phase transitions. Those criteria are the bond orientational correlation function, the Larson-Grier criterion, 2D dynamic Lindemann parameter, the bond-orientational susceptibility, the 2D Hansen-Verlet rule, the L\"{o}wen-Palberg-Simon criterion as well as shape factor and  Minkowski functionals. A very sensitive tool to distinguish different symmetries ist the bond order correlation function $g_6(\mathbf{r})$. The transition from algebraic decay to exponential decay marks the hexatic to isotropic fluid phase transition. The bond order parameter susceptibility provides similar results for the hexatic isotropic transition and might be used as an alternative measure even applicable in poly-crystalline samples \cite{Wang2010} which we may name 'poly-domain hexatic' or 'polyhexaline', since the measure is taken above $T_m$.

The counterpart of $g_6(\mathbf{r})$ is the density-density correlation function (eq.\;\ref{eq_gG}) for the hexatic to crystalline transition where the crossover from quasi-long-range to short-range translational order marks the symmetry breaking temperature. This quantity is rarely used in experiment since the reciprocal lattice vector is not precise to determine if Mermin-Wagner fluctuations are present. Therefore we use a dynamic quantity, the means-squared-displacement with respect to the nearest neighbors, namely the 2D dynamic Lindemann parameter to identify the hexatic to crystalline transition. The 2D Lindemann parameter is a sensitive tool; it stays finite in the crystal but diverges in the fluid phase provided that the system is free of grain boundaries according KTHNY-theory. This should be the case for transitions with continuous character but on the experimental side substrate interaction, density and temperature gradients or large cooling rates may induce grain boundaries.

Since the local order in 2D systems is six-fold in both, the fluid and the solid phase, local measures like the Larson-Grier criterion and the shape factor of voronoi cells do not change significantly crossing transition temperatures and are rather insensitive to global symmetry changes. This is at least true for our system with purely repulsive pair interaction where density differences do not appear in different phases.

The Hansen-Verlet rule modified for two-dimensional systems measures the hight of the first peak of the structure factor. Values between $S(q_0)=4.4$ and $S(q_0)=5.75$ are reported in simulations. In our dipolar system we determined $S(q_0)\simeq 10$ at the melting point. A critical value might be given for individual systems but a universal value should be taken with care. This is the same for the ratio of the long time versus short time diffusion coefficient. In 3D systems the L\"{o}wen-Palberg-Simon criterion states that crystallization takes place at an critical value of $0.1$. In 2D values between 0.072 and 0.099 are found in simulations, whereas in our system we found a value of 0.03. The discrepancies might be explained with grain boundaries where we do not know if those were present in the simulations.

Finally we presented Minkowski functionals as topological measure to identify the phases. Whereas we found Minkowski functional to be sensitive to locally heterogeneous distributions of particles in a binary mixture, they were rather insensitive to global symmetry changes and phase transitions.

\section*{Acknowledgement}

This work was supported by the DFG (Deutsche Forschungsgemeinschaft) in the frame of SFB-TR6 project C2.

\section*{References}
\bibliography{jpcm}{}

\begin{thebibliography}{10}

\bibitem{Chui1983}
S.~T. Chui.
\newblock {\em Phys. Rev. B}, 28(1):178--194, Jul 1983.

\bibitem{Kleinert1983}
H.~Kleinert.
\newblock {\em Phys. Lett. A}, 95(7):381--384, 1983.

\bibitem{Glaser1993}
M.~A. Glaser and N.~A. Clark.
\newblock {\em Adv. Chem. Phys.}, 83:543, 1993.

\bibitem{Lansac2006}
Y.~Lansac, M.~A. Glaser, and N.~A. Clark.
\newblock {\em Phys. Rev. E}, 73:041501, Apr 2006.

\bibitem{Kosterlitz1973}
J.~M. Kosterlitz and D.~J. Thouless.
\newblock {\em Journal of Physics C: Solid State Physics}, 6(7):1181, 1973.

\bibitem{Nelson1979}
D.~R. Nelson and B.~I. Halperin.
\newblock {\em Phys. Rev. B}, 19(5):2457--2484, Mar 1979.

\bibitem{Halperin1978}
B.~I. Halperin and D.~R. Nelson.
\newblock {\em Phys. Rev. Lett.}, 41(2):121--124, Jul 1978.

\bibitem{Young1979}
A.~P. Young.
\newblock {\em Phys. Rev. B}, 19(4):1855--1866, Feb 1979.

\bibitem{Alder1962}
B.~J. Alder and T.~E. Wainwright.
\newblock {{Phase Transition in Elastic Disks}}.
\newblock {\em Phys. Rev.}, 127(2):359--361, Jul 1962.

\bibitem{Mak2006}
C.~H. Mak.
\newblock {\em Phys. Rev. E}, 73:065104, Jun 2006.

\bibitem{Alonso1999}
J.~J. Alonso and J.~F. Fernandez.
\newblock {{van der Waals loops and the melting transition in two dimensions}}.
\newblock {\em Phys. Rev. E}, 59(3):2659--2663, Okt 1999.

\bibitem{Lin2006}
S.~Z. Lin, B.~Zheng, and S.~Trimper.
\newblock {\em Phys. Rev. E}, 73:066106, Jun 2006.

\bibitem{Strandburg1988}
K.~J. Strandburg.
\newblock {\em Rev. Mod. Phys.}, 60(1):161--207, Jan 1988.

\bibitem{Sengupta2000}
S.~Sengupta, P.~Nielaba, and K.~Binder.
\newblock {\em Phys. Rev. E}, 61:6294--6301, Jun 2000.

\bibitem{Binder2002}
K.~Binder, S.~Sengupta, and P.~Nielaba.
\newblock {\em Journal of Physics: Condensed Matter}, 14(9):2323, 2002.

\bibitem{Bernard2011}
E.~P. Bernard and W.~Krauth.
\newblock {\em Phys. Rev. Lett.}, 107:155704, Oct 2011.

\bibitem{Marcus1997}
A.~H. Marcus and S.~A. Rice.
\newblock {\em Phys. Rev. E}, 55(1):637--656, Jan 1997.

\bibitem{Kusner1994}
R.~E. Kusner, J.~A. Mann, J.~Kerins, and A.~J. Dahm.
\newblock {\em Phys. Rev. Lett.}, 73:3113--3116, Dec 1994.

\bibitem{Zahn1999}
K.~Zahn, R.~Lenke, and G.~Maret.
\newblock {\em Phys. Rev. Lett.}, 82(13):2721--2724, Mar 1999.

\bibitem{Zahn2000}
K.~Zahn and G.~Maret.
\newblock {\em Phys. Rev. Lett.}, 85(17):3656--3659, Oct 2000.

\bibitem{Segalman2003}
R.~A. Segalman, A.~Hexemer, R.~C. Hayward, and E.~J. Kramer.
\newblock {\em Macromolecules}, 36(9):3272--3288, 2003.

\bibitem{Angelescu2005}
D.~E. Angelescu, C.~K. Harrison, M.~L. Trawick, R.~A. Register, and P.~M.
  Chaikin.
\newblock {\em Phys. Rev. Lett.}, 95:025702, Jul 2005.

\bibitem{Keim2007}
P.~Keim, G.~Maret, and H.~H. von Gr\"unberg.
\newblock {\em Phys. Rev. E}, 75(3):031402, Mar 2007.

\bibitem{Han2008}
Y.~Han, N.~Y. Ha, A.~M. Alsayed, and A.~G. Yodh.
\newblock {\em Phys. Rev. E}, 77:041406, Apr 2008.

\bibitem{Wang2010}
Z.~Wang, A.~M. Ahmed, A.~G. Yodh, and Y.~Han.
\newblock {\em J. Chem. Phys.}, 132(15):154501--8, April 2010.

\bibitem{Dillmann2008}
P.~Dillmann, G.~Maret, and P.~Keim.
\newblock {\em Journal of Physics: Condensed Matter}, 20(40):404216, 2008.

\bibitem{Peierls1923}
R.~E. Peierls.
\newblock {\em Helv. Phys. Acta}, 7(81), 1923.

\bibitem{Peierls1935}
R.~E. Peierls.
\newblock {\em Ann. l. H. Poincar\'e}, 5(3):177--222, 1935.

\bibitem{Mermin1966}
N.~D. Mermin and H.~Wagner.
\newblock {\em Phys. Rev. Lett.}, 17(22):1133--1136, Nov 1966.

\bibitem{Mermin1968}
N.~D. Mermin.
\newblock {\em Phys. Rev.}, 176(1):250--254, Dec 1968.

\bibitem{Ebert2009}
F.~Ebert, P.~Dillmann, G.~Maret, and P.~Keim.
\newblock {\em Rev. Sci. Instrum.}, 80(8):083902, 2009.

\bibitem{Larsen1996}
A.~E. Larsen and D.~G. Grier.
\newblock {\em Phys. Rev. Lett.}, 76(20):3862--3865, May 1996.

\bibitem{Lindemann1910}
F.~A. Lindemann.
\newblock {\em Phys. Z.}, 609(14):609--612, 1910.

\bibitem{Gilvarry1956}
J.~J. Gilvarry.
\newblock {\em Phys. Rev.}, 102(2):308--316, Apr 1956.

\bibitem{Bedanov1985}
V.~M. Bedanov, G.~V. Gadiyak, and Yu.~E. Lozovik.
\newblock {\em Physics Letters A}, 109(6):289 -- 291, 1985.

\bibitem{Weber1995}
H.~Weber, D.~Marx, and K.~Binder.
\newblock {\em Phys. Rev. B}, 51(20):14636--14651, May 1995.

\bibitem{Hansen1969}
J.-P. Hansen and L.~Verlet.
\newblock {\em Phys. Rev.}, 184(1):151--161, Aug 1969.

\bibitem{Caillol1982}
J.~M. Caillol, D.~Levesque, J.~J. Weis, and J.~P. Hansen.
\newblock {\em Journal of Statistical Physics}, 28(2):325--349, jun 1982.

\bibitem{Ramakrishnan1982}
T.~V. Ramakrishnan.
\newblock {\em Phys. Rev. Lett.}, 48(8):541--545, Feb 1982.

\bibitem{Aeppli1984}
G.~Aeppli and R.~Bruinsma.
\newblock {\em Phys. Rev. Lett.}, 53(22):2133--2136, Nov 1984.

\bibitem{Davey1984}
S.~C. Davey, J.~Budai, J.~W. Goodby, R.~Pindak, and D.~E. Moncton.
\newblock {\em Phys. Rev. Lett.}, 53:2129--2132, Nov 1984.

\bibitem{Lowen1993}
H.~L\"owen, T.~Palberg, and R.~Simon.
\newblock {\em Phys. Rev. Lett.}, 70(10):1557--1560, Mar 1993.

\bibitem{Lowen1996}
H.~L\"owen.
\newblock {\em Phys. Rev. E}, 53(1):R29--R32, Jan 1996.

\bibitem{Zippelius1980}
A.~Zippelius, B.~I. Halperin, and D.~R. Nelson.
\newblock {\em Phys. Rev. B}, 22(5):2514--2541, Sep 1980.

\bibitem{Mouifmmodecheckcelsevcfika2005}
F.~Mou\ifmmode~\check{c}\else \v{c}\fi{}ka and I.~Nezbeda.
\newblock {\em Phys. Rev. Lett.}, 94(4):040601, Feb 2005.

\bibitem{Reis2006}
P.~M. Reis, R.~A. Ingale, and M.~D. Shattuck.
\newblock {\em Phys. Rev. Lett.}, 96(25):258001, Jun 2006.

\bibitem{Mecke1997}
K.~R. Mecke and V.~Sofonea.
\newblock {\em Phys. Rev. E}, 56(4):R3761--R3764, Oct 1997.

\bibitem{C.Beisbart2001}
{C. Beisbart}, {R. Valdarnini}, and {T. Buchert}.
\newblock {\em Astron. Astrophys.}, 379(2):412--425, 2001.

\bibitem{Ebert2009a}
F.~Ebert, G.~Maret, and P.~Keim.
\newblock {\em Eur. Phys. J. E}, 29:311--318, 2009.
\newblock 10.1140/epje/i2009-10490-x.

\bibitem{Hadwiger1957}
H.~Hadwiger.
\newblock {\em Vorlesungen \"uber Inhalt, Oberfl\"ache und Isoperimetrie}.
\newblock Springer Verlag, 1957.

\end{thebibliography}

\end{document}